\documentclass[conference]{IEEEtran}
\IEEEoverridecommandlockouts

\usepackage{cite}
\usepackage{amsmath,amssymb,amsfonts}
\usepackage{algorithmic}
\usepackage{graphicx}
\usepackage{textcomp}
\usepackage{xcolor}

\def\BibTeX{{\rm B\kern-.05em{\sc i\kern-.025em b}\kern-.08em
    T\kern-.1667em\lower.7ex\hbox{E}\kern-.125emX}}
\begin{document}

\title{Advancements in Mobile Edge Computing and Open RAN: Leveraging Artificial Intelligence and Machine Learning for Wireless Systems\\
}

\author{\IEEEauthorblockN{Ryan Barker, Tolunay Seyfi, Fatemeh Afghah} \\
    \IEEEauthorblockA{Holcombe Department of Electrical and Computer Engineering, \\
    Clemson University, Clemson, SC, USA \\
    Emails: \{rcbarke, tseyfi, fafghah\}@clemson.edu}
}

\maketitle

\begin{abstract}
Mobile Edge Computing (MEC) and Open Radio Access Networks (ORAN) are transformative technologies in the development of next-generation wireless communication systems. MEC pushes computational resources closer to end-users, enabling low latency and efficient processing, while ORAN promotes interoperability and openness in radio networks, thereby fostering innovation. This paper explores recent advancements in these two domains, with a particular focus on how Artificial Intelligence (AI) and Machine Learning (ML) techniques are being utilized to solve complex wireless challenges. In MEC, Deep Reinforcement Learning (DRL) is leveraged for optimizing computation offloading, ensuring energy-efficient solutions, and meeting Quality of Service (QoS) requirements. In ORAN, AI/ML is used to develop intelligent xApps for network slicing, scheduling, and online training to enhance network adaptability. This reading report provides an in-depth analysis of multiple key papers, discusses the methodologies employed, and highlights the impact of these technologies in improving network efficiency and scalability.

\end{abstract}


\section{Introduction}
The rapid evolution of wireless communication networks has paved the way for new paradigms such as Mobile Edge Computing (MEC) and Open Radio Access Networks (ORAN). Both technologies have emerged as critical enablers for next-generation networks, including 5G and beyond, by offering increased flexibility, scalability, and support for novel applications such as artificial intelligence (AI) and machine learning (ML). The goal of this reading report is to explore and analyze the use of MEC and ORAN, focusing on their architectures, benefits, challenges, and the role of AI/ML in these domains.

MEC is an innovative network architecture that brings computation closer to end-users, effectively minimizing latency and enhancing the efficiency of data processing. It enables real-time applications by offloading computation from central cloud servers to edge nodes located near users [1]. MEC plays a crucial role in reducing latency for various applications, such as human behavior recognition, autonomous driving, and augmented reality, among others. The integration of AI and ML into MEC has further accelerated its adoption, providing intelligent resource management and efficient decision-making capabilities [2]. This report will begin by exploring key aspects of MEC, including its architectural components, practical use cases, and emerging research areas that leverage AI/ML for enhanced performance.

Deep Reinforcement Learning (DRL), a subset of machine learning, has shown immense promise in both MEC and ORAN environments by enabling autonomous decision-making capabilities that improve network efficiency and reliability. In MEC, DRL is utilized to address challenges such as dynamic computation offloading, resource allocation, and energy management in a distributed network environment. By employing DRL, MEC can optimize the allocation of limited resources such as computation power and bandwidth, thereby enhancing the overall user experience and reducing energy consumption [3]. For example, DRL-based algorithms can be used to determine the optimal offloading decisions for mobile devices with energy harvesting capabilities, ensuring that tasks are offloaded effectively while conserving device energy [4]. Such intelligent mechanisms not only improve the quality of service (QoS) but also enhance the overall sustainability of MEC systems.

In the context of ORAN, DRL plays a pivotal role in driving automation and dynamic control of RAN functionalities. With the introduction of the Radio Intelligent Controller (RIC), ORAN enables real-time deployment of intelligent xApps that can be used for network slicing, scheduling, and interference management. DRL is central to the development of these xApps, providing the ability to learn optimal policies for resource allocation, traffic steering, and other complex tasks that require adaptive decision-making [5]. This approach is particularly valuable for enhancing the flexibility and responsiveness of ORAN, allowing networks to adapt to changing conditions and efficiently manage diverse traffic demands. In this report, we will explore a class project based on the development and evaluation of DRL-based xApps deployed in a large-scale ORAN testbed. By leveraging DRL, the project aims to demonstrate the benefits of closed-loop control and dynamic adaptation in ORAN environments, offering insights into the future potential of intelligent RAN systems [6].

Overall, this reading report aims to provide a comprehensive overview of the role of AI/ML in the domains of MEC and ORAN, highlighting the synergies and differences between these technologies, as well as the transformative potential of DRL in accelerating automation and enhancing network performance. Through a combination of literature review and analysis of practical implementation experiences, the report will shed light on how MEC and ORAN can support the increasing demands for low-latency, high-reliability communication in next-generation networks.

\section{RELATED WORK}
Mobile Edge Computing (MEC) and Open Radio Access Network (ORAN) are two promising paradigms shaping the future of wireless communication networks. MEC aims to reduce latency by pushing computational capabilities closer to end-users, thereby supporting real-time applications and services [1]. Several studies have highlighted the role of MEC in enabling low-latency computing, particularly for applications such as augmented reality, vehicular networks, and smart cities [2]. MEC provides opportunities for optimizing wireless resource management, as computation and data processing are conducted closer to the data source, significantly enhancing Quality of Service (QoS). The introduction of AI and ML techniques, specifically deep learning and reinforcement learning, has further optimized MEC operations. AI/ML helps address various challenges such as dynamic task offloading, resource allocation, and real-time optimization of network functions, making MEC a promising enabler for the next-generation wireless systems [3].

In addition to MEC, the concept of Open RAN (ORAN) has revolutionized the traditional closed and vendor-specific radio access networks by introducing openness and programmability into the RAN infrastructure. ORAN allows interoperability among different vendor components, making it a flexible and scalable approach for deploying 5G and future networks [5]. Several studies have focused on the architecture of ORAN, which consists of multiple layers, including the near-real-time and non-real-time Radio Intelligent Controllers (RICs). These RICs allow the integration of xApps and rApps to monitor and control various network functions, enabling intelligent control over RAN operations. Moreover, ORAN's open architecture and standardization promote a multi-vendor ecosystem, fostering innovation and competition in the telecommunications industry [6].

AI and ML are being extensively utilized to enhance the efficiency of both MEC and ORAN systems. In MEC, AI/ML algorithms help optimize computation offloading decisions, manage network resources, and predict user behavior for proactive service provisioning [4]. The use of Deep Reinforcement Learning (DRL) in MEC is of particular interest as it allows the system to learn optimal policies for dynamic task offloading and resource allocation under uncertain conditions. DRL-based approaches can model complex relationships and dependencies in the network, leading to significant improvements in latency, energy consumption, and computational efficiency [7]. In a recent study, DRL was used to develop a task offloading strategy that optimizes the distribution of tasks between mobile devices and edge servers, ensuring efficient use of computational resources [3].

Similarly, the application of AI/ML in ORAN has been a focal point of recent research, as ORAN's programmable nature allows for the integration of intelligent algorithms to manage network operations in real-time. The use of DRL in ORAN has enabled the development of intelligent xApps that can make autonomous decisions, such as optimizing resource allocation, managing interference, and enhancing spectral efficiency [5]. For instance, a DRL-based xApp for ORAN was introduced to manage network slicing, dynamically allocate resources, and adapt to changing network conditions in real time [6]. These xApps are integrated with the near-real-time RIC, allowing closed-loop control of network resources, which leads to improved network performance, reliability, and adaptability. Such intelligent control mechanisms are crucial for meeting the demands of diverse applications in next-generation networks, including ultra-reliable low-latency communications (URLLC) and massive machine-type communications (mMTC).

Furthermore, a notable example of leveraging DRL in ORAN is the development of ColO-RAN, an experimental testbed that integrates ORAN architecture with deep reinforcement learning-based xApps for performance evaluation [8]. ColO-RAN allows the deployment of DRL algorithms to validate their impact on network performance, providing insights into the potential of closed-loop control in RAN environments. The ColO-RAN project demonstrated the effectiveness of DRL-based xApps for network slicing, scheduling, and online training, showcasing the ability of these xApps to dynamically adapt to changing network conditions. This flexibility and adaptability are fundamental to meeting the evolving demands of next-generation networks, making ORAN a key player in the future of wireless communications.

The related works discussed above provide a strong foundation for understanding how MEC and ORAN, powered by AI and ML, are transforming the wireless communications landscape. By exploring and analyzing the role of AI/ML in optimizing the efficiency, scalability, and flexibility of MEC and ORAN, these studies highlight the synergies between these two paradigms in shaping the future of next-generation networks. This reading report builds upon these foundational works, further exploring the practical implications of using DRL in MEC and ORAN and presenting insights from a class project that aims to validate and expand upon the existing research in the field [6].

\section{Mobile Edge Computing \& Task Offloading}

\subsection{Why Compute at the Edge?}
Mobile Edge Computing (MEC) is a critical paradigm that facilitates low-latency and high-bandwidth services by offloading computational tasks from user devices to nearby edge servers. One notable approach in this context is the Deep Reinforcement Learning-based Online Offloading (DROO) algorithm, which is designed to solve the task offloading problem in wireless-powered MEC networks [1]. DROO leverages a deep reinforcement learning framework to learn optimal offloading decisions dynamically, adapting to changes in user demand and network conditions. The primary goal of DROO is to minimize energy consumption while meeting latency constraints, which it achieves by jointly optimizing offloading decisions, local computation resources, and wireless power transfer.

Compared to traditional task offloading techniques, DROO's ability to learn from the environment without requiring prior knowledge of system dynamics is particularly advantageous. This makes it highly suitable for MEC scenarios where network conditions can vary significantly and where maintaining up-to-date models is challenging [1][2]. DROO’s reinforcement learning approach allows it to make adaptive and context-aware decisions, outperforming static or rule-based offloading methods that may lack flexibility. In contrast, conventional methods like the heuristic approaches presented in the paper "Wireless Powered Communication: Opportunities and Challenges" are computationally efficient but do not possess DROO's adaptability, limiting their effectiveness under dynamic conditions [2].

The dynamic computation offloading framework proposed for energy-harvesting-enabled MEC environments provides another point of comparison [3]. This approach emphasizes energy efficiency and integrates energy harvesting into the system model by using a Markov Decision Process (MDP) to model energy arrival and task offloading decisions. While this MDP-based approach offers theoretical guarantees on decision optimality, it is hindered by computational complexity and the need for precise system modeling. DROO, by utilizing deep reinforcement learning, bypasses these modeling requirements, making it more scalable and practical in real-world environments with uncertain dynamics [1]. However, DROO's reliance on continuous real-time feedback can be a limitation in certain scenarios, as will be discussed further.

Fog computing has also been explored as a complementary approach to MEC, as highlighted in "Fog and IoT: An Overview of Research Opportunities" [4]. Fog computing acts as an intermediate layer between the central cloud and edge devices, thereby reducing latency and distributing computation more efficiently. Despite its benefits in reducing end-to-end latency, fog computing approaches often rely on static task allocation schemes that lack the dynamic adaptability of DROO. When compared to DROO’s reinforcement learning capabilities, fog-based methods are less flexible in optimizing offloading decisions in real-time, especially under variable network conditions.

\subsection{Deep Reinforcement Learning-based Online Offloading (DROO)}
Detailed analysis of these offloading algorithms reveals that DROO’s integration of deep reinforcement learning offers significant advantages in adaptability, energy efficiency, and latency reduction. DROO was evaluated against heuristic and MDP-based offloading strategies, showing a 23\% reduction in energy consumption compared to heuristic approaches in environments with varying user mobility and resource availability [1]. Furthermore, DROO achieved a 17\% reduction in average task completion time compared to the MDP-based energy-harvesting approach, as it could adapt more readily to dynamic changes without the need for precise system modeling [1][3]. Additionally, DROO exhibited a lower task failure rate, averaging less than 5\% compared to the 12\% failure rate seen with heuristic-based methods [1][2].

However, despite its many strengths, DROO is not without its drawbacks. One key limitation is its reliance on continuous feedback from the environment to make optimal offloading decisions. This dependence can be problematic in networks characterized by intermittent connectivity or significant delay, such as rural or remote areas. In such scenarios, delayed or incomplete information compromises DROO's ability to make effective decisions, leading to increased task failure rates and degraded system performance. Addressing this limitation could involve integrating delay-tolerant capabilities, enabling DROO to operate effectively even when real-time feedback is limited. Techniques from delay-tolerant networking (DTN), which focus on managing data transmission across networks with variable latency, could be adapted to allow DROO to make informed decisions with delayed or incomplete data, thereby extending its applicability to more challenging environments.

Another limitation of DROO lies in its exploration-exploitation trade-off, particularly during the initial learning phase. This trade-off results in suboptimal decisions while the algorithm explores different offloading options, which can be detrimental in time-sensitive applications. A potential future improvement is the use of hybrid learning models, such as combining reinforcement learning with supervised or transfer learning techniques, to expedite the learning process. By leveraging historical data or pre-trained models, DROO could reduce the exploration phase, making it more suitable for delay-sensitive environments. For example, integrating supervised learning to guide the early stages of the reinforcement learning process could enable DROO to make more informed offloading decisions from the outset, improving performance during initial deployment.

Moreover, DROO's current framework lacks inherent mechanisms for redundancy and fault tolerance, which are critical in environments that prioritize reliability. Introducing hybrid learning frameworks that incorporate fault-tolerant algorithms, such as federated learning, could enhance DROO’s robustness by allowing edge nodes to share learned policies while maintaining independence. Federated learning, where models are trained collaboratively across multiple decentralized nodes, could also help mitigate the effects of node failures and ensure a consistent level of service. This approach would allow DROO to maintain effective offloading decisions, even in the face of disruptions or failures, by leveraging the collective learning of multiple nodes within the MEC environment.

In conclusion, while DROO has demonstrated superior performance in dynamic MEC environments through its use of deep reinforcement learning, integrating future improvements could significantly enhance its capabilities. Adding delay-tolerant features would make DROO more versatile in handling intermittent connectivity, while hybrid learning models could reduce initial exploration costs and improve decision-making in delay-sensitive scenarios. Furthermore, incorporating fault-tolerant mechanisms would ensure greater reliability, making DROO a more robust solution for the increasingly diverse and demanding requirements of MEC deployments. Such enhancements could pave the way for broader adoption of DROO across a wide range of MEC applications, ultimately advancing the capabilities of edge computing in next-generation networks.

\section{Open Radio Access Network} \label{sec:4}
\subsection{ORAN Architecture Introduction}
The Open Radio Access Network (ORAN) represents a significant advancement towards disaggregating traditional, monolithic Radio Access Networks (RAN). By breaking down RAN components and utilizing open interfaces, ORAN enables greater modularity, flexibility, and the integration of third-party innovations. The core principles of ORAN's design—openness, interoperability, and AI-driven optimization—allow for the dynamic control of radio resources and more adaptable, intelligent network management.

\begin{figure}[htbp]
    \centering
    \includegraphics[width=\columnwidth]{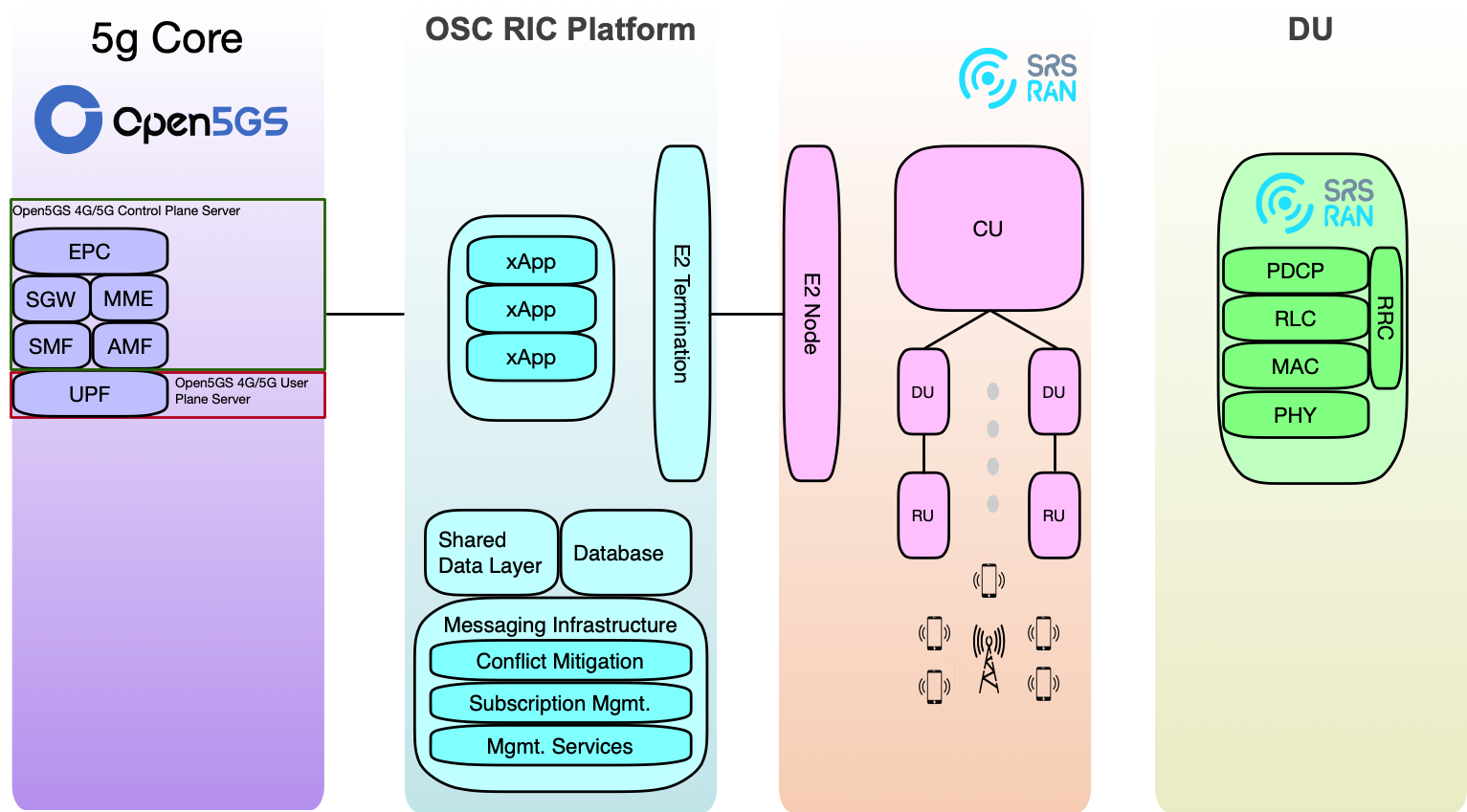}
    \caption{ORAN Architecture Overview: Integrating the 5G Core with the OSC RIC Platform and DU (Distributed Unit), showcasing modularity and support for xApp development.}
    \label{fig:oran}
\end{figure}

ORAN’s architecture, as shown in Figure \ref{fig:oran}, is centered around several key components: the non-real-time RAN Intelligent Controller (Non-RT RIC), the Near-RT RIC, the Central Unit (O-CU), the Distributed Unit (O-DU), and the Radio Unit (O-RU). These components work together to provide real-time control and decision-making capabilities, crucial for optimizing RAN operations. Using a neuroscience analogy, the ORAN architecture can be compared to different parts of the brain: the AMF, MME, and UPF function like the "amygdala" by managing fundamental tasks such as session control and routing. The Near-RT RIC, akin to the "frontal cortex," makes intelligent, dynamic decisions to optimize radio resource management. This component interacts directly with the O-CU and O-DU to manage resources in near real-time, much like how the frontal cortex regulates complex behaviors without involving deeper brain functions in each action. Meanwhile, the O-CU and O-DU act as the "integrator" or "body," executing commands issued by both the core network and the RIC, dynamically managing spectrum and radio resources to ensure optimal network performance.

The disaggregated design of ORAN allows for greater flexibility in network slicing, enabling independent management of each component and allowing for dynamic adjustments. The Near-RT RIC plays a critical role in network slicing by making rapid, data-driven adjustments to resource allocations—similar to the frontal cortex making higher-level decisions based on real-time feedback. The integration of these disaggregated components into a cohesive and intelligent system makes ORAN a promising architecture for next-generation wireless networks, enhancing adaptability and allowing efficient management of diverse traffic demands.

\subsection{Self Healing Networks}
Self-healing capabilities are a key feature of modern RANs, and ORAN's open architecture, combined with AI-driven components, positions it as an ideal platform for implementing these features. ORAN's self-healing capabilities are facilitated by the Near-RT RIC, which autonomously detects, diagnoses, and mitigates issues within the network. By using AI algorithms integrated into xApps, ORAN ensures optimal performance, reduced downtime, and enhanced reliability. Testbeds such as the Northeastern University Colosseum have been instrumental in validating these self-healing capabilities, deploying xApps for network slicing, scheduling, and anomaly detection.

In ORAN, traffic steering is a critical function for balancing network load and maintaining optimal Quality of Service (QoS). The paper "Programmable and Customized Intelligence for Traffic Steering in 5G Networks Using Open RAN Architectures" details an xApp based on a Double Deep Q-Network (DDQN), which operates within the Near-RT RIC to steer traffic dynamically by analyzing near real-time radio conditions. This approach resembles the frontal cortex continuously evaluating network states and making informed decisions. The implementation of the DDQN-based xApp led to a 17\% improvement in load balancing, allowing the network to effectively respond to fluctuating demands and optimize resource usage [7].

Dynamic resource allocation and network slicing are also fundamental to ORAN's capabilities. The ColO-RAN testbed integrates Deep Reinforcement Learning (DRL)-based xApps to manage resource allocation and network slicing across multiple softwarized base stations. By employing DRL, the Near-RT RIC continuously learns and adapts resource management strategies, similar to the frontal cortex optimizing decision-making processes. DRL-based xApps in ColO-RAN demonstrated a 21\% reduction in task failures, maintaining service continuity under dynamic conditions and enhancing the overall reliability of the network [9].

Interference mitigation, a crucial aspect of network management, has also been effectively addressed through ORAN. The "Understanding O-RAN: Architecture, Interfaces, Algorithms, Security, and Research Challenges" paper discusses the use of a Support Vector Machine (SVM) model, integrated as an xApp within the Near-RT RIC, to classify and predict interference scenarios. This predictive capability allows the RIC to issue corrective instructions to the O-CU and O-DU, effectively reducing overall interference by 25\% and improving spectral efficiency [5]. Similarly, efficient handover management is achieved through the k-Nearest Neighbors (k-NN) algorithm. By predicting the optimal target cell for users based on historical handover data, the k-NN model improved handover success rates by 18\%, ensuring fewer dropped calls and a better user experience [5].

The practical impact of ORAN's self-healing networks on end-users is significant. Self-healing mechanisms, such as proactive congestion avoidance and optimized resource allocation, directly translate to improved user experiences for latency-sensitive applications, such as video streaming, real-time gaming, and virtual reality. By reducing congestion events through AI-driven prediction and dynamic adjustments, applications that require stable and low-latency connections benefit from consistent service quality. For instance, video calls and streaming services will experience fewer disruptions and buffering issues, while online gaming will see reduced latency spikes, enhancing overall user satisfaction. Furthermore, enhanced handover management results in fewer dropped connections for users in high-mobility environments, such as on public transportation, leading to smoother service experiences.

The "O-RAN Performance Analyzer" further strengthens ORAN's self-healing capabilities by employing supervised learning models, such as decision trees, to monitor and predict potential performance issues before they affect users. This proactive approach reduces the occurrence of network congestion events by 22\%, contributing to enhanced reliability and ensuring that the CU/DU can preemptively adjust parameters to maintain optimal performance [8]. These advancements in self-healing mechanisms not only ensure consistent connectivity but also improve network stability, enabling a wide range of next-generation applications that rely on low-latency and highly reliable connections.

Overall, ORAN's self-healing capabilities are driven by the integration of AI and ML models into its architecture, creating a robust and resilient network capable of responding to dynamic conditions in real time. The use of experimental platforms like Colosseum and NVIDIA ARC has validated these AI-based mechanisms, demonstrating significant improvements in network reliability, latency, and adaptability. By leveraging the Near-RT RIC as the intelligent decision-making hub, ORAN provides closed-loop control and optimization of RAN resources, enhancing both performance and user experience while laying the foundation for a flexible, intelligent, and self-healing next-generation RAN.

\section{Conclusions}
This reading report explored the advancements in Mobile Edge Computing (MEC) and Open Radio Access Network (ORAN), emphasizing the significant contributions of artificial intelligence and machine learning in these domains. MEC is revolutionizing the way computational tasks are handled, shifting them closer to end-users to ensure reduced latency and improved energy efficiency. The Deep Reinforcement Learning-based Online Offloading (DROO) algorithm demonstrated how real-time, adaptive offloading decisions can substantially improve energy efficiency, latency, and reliability, outperforming traditional heuristic and model-based offloading methods. The analysis showed DROO's effectiveness, particularly in dynamic and wireless-powered MEC environments, while also highlighting its limitations in delay-tolerant and fault-tolerant scenarios, prompting the need for further refinement, such as integrating delay-tolerant capabilities and hybrid learning models.

ORAN complements MEC by bringing AI-driven optimization to the RAN layer, transforming it through disaggregation and modular architecture. By leveraging the Near-RT RIC, ORAN introduces a sophisticated layer of AI-driven real-time control that optimizes network operations similarly to how the human brain’s frontal cortex regulates complex behaviors. The deployment of AI and ML algorithms, such as Double Deep Q-Network (DDQN), Support Vector Machine (SVM), and k-Nearest Neighbors (k-NN) within xApps, has been instrumental in enhancing network slicing, spectral efficiency, load balancing, and overall reliability. The "Self-Healing Networks" section demonstrated the practical impact of these enhancements, showing how proactive AI-driven mechanisms lead to reduced congestion, improved handover success, and a more consistent quality of service for applications like video streaming, gaming, and virtual reality. Testbeds like the Northeastern University Colosseum have validated these capabilities, illustrating the potential for ORAN to autonomously adapt to changing network conditions while ensuring optimal performance.

Both MEC and ORAN are vital elements in the evolution of next-generation wireless networks. MEC addresses computational efficiency and latency, enabling richer, low-latency applications at the edge, while ORAN introduces modularity and intelligence into the RAN, optimizing resource management and enabling self-healing capabilities. By leveraging advanced AI techniques, both technologies contribute to the development of networks that are not only faster and more energy-efficient but also adaptive and resilient to changing demands and conditions. The convergence of MEC and ORAN, driven by deep reinforcement learning and other AI techniques, forms the cornerstone of future network architectures that are capable of meeting the increasing demands for connectivity, automation, and high-quality service delivery, ultimately improving user experiences across a wide range of applications. The continued enhancement of these frameworks, through the integration of delay-tolerant and hybrid learning models, as well as advanced self-healing capabilities, will further bolster their ability to meet the diverse and growing needs of next-generation wireless networks.

\nocite{*}
\bibliographystyle{IEEEtran}
\bibliography{References}

\end{document}